\pdfoutput=1

\documentclass[11pt]{article}

\usepackage{ACL2023}

\usepackage[hang,flushmargin]{footmisc}

\usepackage{times}
\usepackage{latexsym}

\usepackage[T1]{fontenc}

\usepackage[utf8]{inputenc}

\usepackage{microtype}

\usepackage{inconsolata}

\usepackage{booktabs}
\usepackage{multirow}
\usepackage{graphicx}
\usepackage{subcaption}
\usepackage{mdframed}
\usepackage{amsmath,amsfonts,bm}
 \usepackage{enumitem}
 \usepackage[official]{eurosym}

\newcommand{\miracl}[0]{MIRACL}
\newcommand{\beir}[0]{BEIR}
\newcommand{\cohere}[0]{Cohere}
\newcommand{\colarge}[0]{{\cohere$_\text{\texttt{large}}$}}
\newcommand{\cosmall}[0]{{\cohere$_\text{\texttt{small}}$}}
\newcommand{\openai}[0]{OpenAI}
\newcommand{\aalph}[0]{Aleph-Alpha}

\newcommand\Ar{Arabic}
\newcommand\Bn{Bengali}

\newcommand\Es{Spanish}
\newcommand\Fa{Persian}
\newcommand\Fi{Finnish}
\newcommand\Fr{French}
\newcommand\Hi{Hindi}
\newcommand\Id{Indonesian}
\newcommand\Ja{Japanese}
\newcommand\Ko{Korean}
\newcommand\Ru{Russian}
\newcommand\Sw{Swahili}
\newcommand\Te{Telugu}
\newcommand\Th{Thai}
\newcommand\Zh{Chinese}
\newcommand\De{German}
\newcommand\Yo{Yoruba}

\title{Evaluating Embedding APIs for Information Retrieval}

\author{Ehsan Kamalloo$^{\dagger}$ \quad Xinyu Zhang$^{\dagger}$ \quad Odunayo Ogundepo$^{\dagger}$ \quad Nandan Thakur$^{\dagger}$ \\ {\bf David Alfonso-Hermelo}$^{\S}$ \quad {\bf Mehdi Rezagholizadeh}$^{\S}$ \quad {\bf Jimmy Lin}$^{\dagger}$ \\[1ex]
  $^{\dagger}$ David R. Cheriton School of Computer Science, University of Waterloo \\ $^{\S}$ Huawei Noah's Ark Lab \\[1ex]
  \texttt{ekamalloo@uwaterloo.ca} \\
  }

\begin{document}
\maketitle
\begin{abstract}
The ever-increasing size of language models curtails their widespread availability to the community, thereby galvanizing many companies into offering access to large language models through APIs.
One particular type, suitable for dense retrieval, is a semantic embedding service that builds vector representations of input text.
With a growing number of publicly available APIs, our goal in this paper is to analyze existing offerings in realistic retrieval scenarios, to assist practitioners and researchers in finding suitable services according to their needs.
Specifically, we investigate the capabilities of existing semantic embedding APIs on domain generalization and multilingual retrieval.
For this purpose, we evaluate these services on two standard benchmarks, BEIR and MIRACL.
We find that re-ranking BM25 results using the APIs is a budget-friendly approach and is most effective in English, in contrast to the standard practice of employing them as first-stage retrievers.
For non-English retrieval, re-ranking still improves the results, but a hybrid model with BM25 works best, albeit at a higher cost.
We hope our work lays the groundwork for evaluating semantic embedding APIs that are critical in search and more broadly, for information access.
\end{abstract}

\section{Introduction}

Language models (LMs), pre-trained on a massive amount of text, power dense retrieval models in ad hoc retrieval \cite{lin2020pretrained}.
Dense retrievers (\citealt{lee-etal-2019-latent,karpukhin-etal-2020-dense,ance,khattab2020colbert,hofstatter2021efficiently,contriever}; {\em inter alia}) essentially measure relevance via similarity between the representations of documents and queries.
As LMs are rapidly scaling up to gigantic models (\citealt{gpt2,gpt3,jurassic1,palm,tnlg}, \textit{inter alia}), their use as the backbone of dense retrieval models has become limited primarily because large language models (LLMs) are computationally expensive and deploying them on commodity hardware is cumbersome and often impractical.

To alleviate this problem, many companies, e.g., {\openai}, and {\cohere}, set out to offer access to their proprietary LLMs through a family of APIs.
For dense retrieval, semantic embedding APIs are designed to provide LLM representations for queries as well as documents.
These APIs are especially appealing in the IR ecosystem because they afford practitioners and researchers the benefit of scale and allow for wider outreach of LLMs in IR.
However, although nowadays, the surge of companies offering such APIs with various model sizes has given us more options, a lack of thorough analysis of these APIs has made it more difficult to determine one's best option for a particular use-case.
Besides, LLM-based APIs are often expensive and experimenting with all of them to determine the most suitable is prohibitively costly. 

In this paper, we analyze embedding APIs for various realistic scenarios in ad hoc retrieval.
To this end, we select three embedding APIs available on the market, i.e., {\openai}, {\cohere}, and {\aalph}, and assess their usability and effectiveness on two crucial directions that stand at the core of most IR applications. 

First, we study domain generalization where retrieval is conducted over collections drawn from a broad range of domains.
Understanding for which domains embedding APIs work well or poorly elucidates their limitations while setting the stage for their wide adoption in their successful domains.
We leverage the widely adopted BEIR benchmark \cite{beir} for this purpose.
On BEIR, we use the APIs as re-rankers on top of BM25 retrieved documents because the large size of document collections in BEIR makes full-ranking (i.e., first-stage retrieval) via the APIs impractical.
Our results show that embedding APIs are reasonably effective re-rankers in most domains, suggesting that re-ranking is not only budget-friendly, but also is effective.
However, we find that on datasets collected via lexical matching, they struggle.
In particular, BM25 outperforms the full-fledged embedding APIs on BioASQ (bio-medical retrieval) and Signal1M (tweet retrieval). 

We also explore the capabilities of embedding APIs in multilingual retrieval where they are tested in several non-English languages, ranging from low-resource to high-resource languages.
More precisely, we use MIRACL \cite{miracl}, a large-scale multilingual retrieval benchmark that spans 18 diverse languages.
The manageable size of the corpora allow us to evaluate the APIs as full-rankers as well as re-rankers.
We find that the winning recipe for non-English retrieval is not re-ranking, unlike retrieval on English documents.
Instead, building hybrid models with BM25 yields the best results.
Our experiments also indicate that the APIs are powerful for low-resource languages, whereas on high-resource languages, open-source models work better.

Overall, our findings offer insights on using embedding APIs in real-world scenarios through two crucial aspects of IR systems.
In summary, our key contributions are:

\begin{itemize}[leftmargin=*]
    \item We extensively review the usability of commercial embedding APIs for realistic IR applications involving domain generalization and multilingual retrieval.
    \item We provide insights on how to effectively use these APIs in practice.
\end{itemize}

\noindent We hope our work lays the groundwork for thoroughly evaluating APIs that are critical in search and more broadly, for information access.

\section{Related Work}

\paragraph{Sentence Embeddings.}
Numerous studies have attempted to build universal representations of sentences using supervision via convolutional neural networks
\cite{kalchbrenner-etal-2014-convolutional}, recurrent neural networks
\cite{conneau-etal-2017-supervised}, or Transformers \cite{cer-etal-2018-universal}.
Other approaches learn sentence embeddings in a self-supervised fashion \cite{kiros2015skip} or in an unsupervised manner \cite{zhang-etal-2020-unsupervised,li-etal-2020-sentence}.
Recent techniques frame the task as a contrastive learning problem \cite{reimers-gurevych-2019-sentence,li-etal-2020-sentence,gao-etal-2021-simcse,kim-etal-2021-self,ni-etal-2022-sentence}.
Embedding APIs largely follow a similar strategy to generate sentence representations \cite{neelakantan2022text}.

\paragraph{Dense Retrieval.} 
While the paradigm has been around for a long time \cite{yih-etal-2011-learning}, the emergence of pre-trained LMs brought dense retrieval \cite{lee-etal-2019-latent, karpukhin-etal-2020-dense} to the mainstream in IR.
Recent dense retrieval models adopt a bi-encoder architecture and generally use contrastive learning to distinguish relevant documents from non-relevant ones \cite{lin2020pretrained}, similar to sentence embedding models. 
LMs are shown to be an effective source to extract representations \cite{karpukhin-etal-2020-dense, ance, hofstatter2021efficiently, khattab2020colbert, izacard2021distilling, contriever}.
This essentially means that with LMs as the backbones and analogous objectives, dense retrievers and sentence embedding models have become indistinguishable in practice.

\section{APIs}

Semantic embedding APIs are generally based on the so-called \emph{bi-encoder} architecture, where queries and documents are fed to a fine-tuned LM in parallel \cite{seo-etal-2018-phrase,karpukhin-etal-2020-dense,reimers-gurevych-2019-sentence}.
The key ingredient of bi-encoders is contrastive learning, whose objective is to enable models to distinguish relevant documents from non-relevant ones.
In our experiments, we adopt the following semantic embedding APIs, presented alphabetically:\footnote{Information about the number of parameters is obtained from \url{https://crfm.stanford.edu/helm/latest/}, accessed on May 15, 2023.}

\paragraph{{\aalph}:} This company has trained a family of multilingual LMs, named \texttt{luminous},\footnote{\url{https://docs.aleph-alpha.com/docs/introduction/luminous/}} with three flavours in size, \texttt{base} (13B), \texttt{extended} (30B), and \texttt{supreme} (70B).
The \texttt{luminous} models support five high-resource languages:\ English, French, German, Italian, and Spanish. 
However, no information is available about the data on which these LMs are trained. 
We used \texttt{luminous}$_\text{\texttt{base}}$ that  projects text into 5120-dimension embedding vectors.

\paragraph{{\cohere}:} This company offers LMs for producing semantic representations in two sizes: \texttt{small} (410M) and \texttt{large} (6B), generating 1024-dimension and 4096-dimension embedding vectors, respectively.
Models are accompanied by model cards \cite{mitchell2019model}.\footnote{\url{https://docs.cohere.ai/docs/representation-card}} Cohere also provides a multilingual model, \texttt{multilingual-22-12},\footnote{\url{https://txt.cohere.ai/multilingual/}} that is trained on a large multilingual collection comprising 100+ languages.
The data consists of 1.4 billion question/answer pairs mined from the web.
The multilingual model maps text into 768-dimension embedding vectors.

\paragraph{{\openai}:}
The company behind the GPT models \cite{gpt2,gpt3,instructgpt} also offers an embedding service.
We use the recommended second-generation model, \texttt{text-embedding-ada-002} \cite{neelakantan2022text} that embeds text into a vector of 1536 dimensions.
The model, initialized from a pre-trained GPT model, is fine-tuned on naturally occurring paired data with no explicit labels, mainly scraped from the web, using contrastive learning with in-batch negatives.

\medskip
\noindent All the APIs described above use Transformer-based language models \cite{vaswani2017attention}, but differ from each other in various ways:

\begin{itemize}[leftmargin=*]
    \item \textbf{Model architecture:} The companies built their models in different sizes, with differences in the number of hidden layers, number of attention heads, the dimension of output layers, etc.
    Other subtle differences in the Transformer architecture are also likely, e.g., where to apply layer normalization in a Transformer layer \cite{xiong2020layer}.
    Additional differences lie in the vocabulary because of different tokenization methods in the pre-trained LM that was used to initialize these embedding models for fine-tuning.

    \item \textbf{Training:} While contrastive learning is at the core of these models, they may vary substantially in details, e.g., the contrastive learning objective and negative sampling strategies.
    Also, the choice of hyper-parameters such as the number of training steps, learning rate, and optimizer is another key difference.

    \item \textbf{Data:} Chief among the differences is the data on which the embedding models are trained.
    As OpenAI and Cohere state in their documentation, the data is mostly mined from the web, but the details of the data curation process remain largely unknown.
    In addition, considering that each company has its own models, differences in pre-training corpora form yet another important variable in the complex process of building embedding APIs.

\end{itemize}

\noindent
These distinctions may potentially lead to substantial differences in the overall effectiveness of the embedding APIs.
Nevertheless, due to the non-disclosure of several details by the API providers, it remains challenging to identify the specific factors that contribute to the strengths and weaknesses of embedding models.
Yet, as the number of such APIs continues to grow, we believe that high-level comparisons on standard benchmarks can provide valuable insights into how well these models operate under various practical scenarios.
For practitioners building systems on top of these APIs, this comparative analysis is useful as they are primarily interested in the end-to-end effectiveness and performance of these APIs and are not typically concerned with their minutiae.

\subsection{Usability}

One of the critical advantages of using embedding APIs is their ease-of-use.
For IR applications, even running an LLM to encode large document collections requires hefty resources, let alone training retrieval models.
Thus, the emergence of such APIs makes LLMs more accessible to the community and paves the way for faster development of IR systems---this is most definitely a positive development.
However, these advantages rest on the usability of the APIs.
In this section, we briefly overview some factors that affect the usability of embedding APIs.

\paragraph{Setup.} 
Basic information on how users can set up the proper environment to use the embedding APIs is the first step.
All three companies provide detailed introductory documentation for this purpose.
The procedure is nearly identical for all three at a high level:\ users need to create an account and generate an API key for authentication.
The companies also furnish web interfaces that enable users to monitor their usage history and available credit, in addition to configuring limits to prevent unintended charges.

\paragraph{Client libraries.}
All three companies have developed open-source client libraries to facilitate access to the APIs.
{\openai} provides libraries for Python and Node.js; there are also unofficial community libraries for other programming languages.
{\cohere} offers development toolkits in Python, Node.js, and Go.
{\aalph} provides a library in Python.

All libraries are structured in a similar way.
One difference we notice is that {\cohere} has a text truncation feature when the input text exceeds the API's input length limit. 
{\openai} and {\aalph} raise an error in this case, meaning that API users need to implement additional checks to avoid such exceptions.
On the other hand, {\cohere}'s API can truncate text from the left or the right, and can also provide an average embedding for long texts up to 4096 tokens by averaging over 512-token spans.

\paragraph{Documentation.} 
All three companies provide a technical API reference, explaining inputs, responses, and errors of their APIs. 
Additionally, all companies provide tutorials and blog posts with examples on how to use their client libraries.

\paragraph{Latency.} The APIs are all offered with a liberal rate limit, i.e., {\openai} at 3K requests per minute, and {\cohere} at 10K requests per minute.\footnote{We were not able to find the rate limits for {\aalph}.} 
We find that API calls are mostly reliable and request service errors are scattershot.
Each API call takes up to roughly 400ms, consistent across all three companies (at least at the time of our experiments).
However, latency presumably depends on the server workload and other factors because we observe variability at different points in time.

We also find that latency depends on the input length, as computing embeddings for queries is generally faster than computing embeddings for documents (as expected).
Finally, we appreciate that {\cohere}'s and {\openai}'s APIs support bulk calls of up to 96 and 2048 texts per call, respectively, whereas for {\aalph}, only one text can be passed in each API call.
This bulk call feature considerably speeds up encoding document collections.

\paragraph{Cost.} 
Our analysis is based on information reported as of Feb 1, 2023.
{\openai} and {\aalph} charge based on the number of tokens and model size: \texttt{ada2} and luminous$_\text{\texttt{base}}$ cost \$0.0004 USD and \euro{}0.078 $\approx$ \$0.086\footnote{\euro{}1.00 $\sim$ \$1.10 as of Feb 1, 2023} per 1,000 tokens.
On the other hand, {\cohere} follows a simpler cost structure, charging based only on the number of API calls, i.e., \$1.00 USD per 1,000 calls.
Our re-ranking experiments on BEIR cost around \$170 USD on {\openai}, whereas it would cost roughly \$2,500 USD on {\cohere} based on their quoted prices.
The cost of our re-ranking experiments on MIRACL for three languages ({\De}, {\Es}, and {\Fr}) hovers around \euro{}116 $\approx$ \$128 using {\aalph} and {\cohere}. 
{\cohere} offers a free-tier access with a restricted API call rate limit of 100 calls per minute, which we opted for, albeit sacrificing speed.

\section{Experiments}

In this section, our main goal is to evaluate embedding APIs in two real-world scenarios that often arise in IR applications: domain generalization and multilingual retrieval.

\begin{table*}[t!]
\centering
\small
\resizebox{\textwidth}{!}{ 
\begin{tabular}{llccccccc}
\toprule
\multirow{2}{*}{\textbf{Task}}& \multirow{2}{*}{\textbf{Domain}}& \multicolumn{3}{c}{\textit{Full-ranking}}& \multicolumn{4}{c}{\textit{BM25 Top-100 Re-rank}} \\
 & & \textbf{BM25}& \textbf{TASB}& \textbf{cpt-S}& \textbf{TASB}& \textbf{{\cohere}}$_\text{\texttt{large}}$& \textbf{{\cohere}}$_\text{\texttt{small}}$& \textbf{{\openai}}$_\text{\texttt{ada2}}$ \\
\cmidrule(lr){1-2} \cmidrule(lr){3-5} \cmidrule(lr){6-9} 
TREC-COVID& \scriptsize{Bio-Medical}& 0.595& 0.319& 0.679& 0.728& 0.801& 0.776& \textbf{0.813}\\
BioASQ& \scriptsize{Bio-Medical}& \textbf{0.523}& 0.481& -& 0.467& 0.419& 0.423& 0.491\\
NFCorpus& \scriptsize{Bio-Medical}& 0.322& \textbf{0.360}& 0.332& 0.334& 0.347& 0.324& 0.358\\
NQ& \scriptsize{Wikipedia}& 0.306& 0.463& -& 0.452& \textbf{0.491}& 0.453& 0.482\\
HotpotQA& \scriptsize{Wikipedia}& 0.633& 0.584& 0.594& 0.628& 0.580& 0.523& \textbf{0.654}\\
FiQA-2018& \scriptsize{Finance}& 0.236& 0.300& 0.384& 0.308& \textbf{0.411}& 0.374& \textbf{0.411}\\
Signal-1M& \scriptsize{Twitter}& \textbf{0.330}& 0.288& -& 0.329& 0.306& 0.295& 0.329\\
TREC-NEWS& \scriptsize{News}& 0.395& 0.377& -& 0.436& 0.461& 0.447& \textbf{0.495}\\
Robust04& \scriptsize{News}& 0.407& 0.428& -& 0.399& 0.489& 0.467& \textbf{0.509}\\
ArguAna& \scriptsize{Misc.}& 0.397& 0.427& 0.470& 0.436& 0.467& 0.438& \textbf{0.567}\\
Touché-2020& \scriptsize{Misc.}& \textbf{0.442}& 0.163& 0.285& 0.292& 0.276& 0.275& 0.280\\
CQADupStack& \scriptsize{StackEx.}& 0.302& 0.314& -& 0.324& \textbf{0.411}& 0.384& 0.391\\
Quora& \scriptsize{Quora}& 0.789& 0.835& 0.706& 0.841& \textbf{0.886}& 0.866& 0.876\\
DBPedia& \scriptsize{Wikipedia}& 0.318& 0.384& 0.362& 0.389& 0.372& 0.344& \textbf{0.402}\\
SCIDOCS& \scriptsize{Scientific}& 0.149& 0.149& -& 0.156& \textbf{0.194}& 0.182& 0.186\\
FEVER& \scriptsize{Wikipedia}& 0.651& 0.700& 0.721& 0.728& 0.674& 0.617& \textbf{0.773}\\
Climate-FEVER& \scriptsize{Wikipedia}& 0.165& 0.228& 0.185& 0.243& \textbf{0.259}& 0.246& 0.237\\
SciFact& \scriptsize{Scientific}& 0.679& 0.643& 0.672& 0.661& 0.721& 0.670& \textbf{0.736}\\
\cmidrule(lr){1-2} \cmidrule(lr){3-5} \cmidrule(lr){6-9} 
\multicolumn{2}{c}{Avg. nDCG@10}& 0.424& 0.414& -& 0.453& 0.476& 0.450& \textbf{0.500}\\
\bottomrule
\end{tabular}
}
\caption{Results (nDCG@10) on the {\beir} benchmark for full-ranking and BM25 re-ranking experiments. \textbf{cpt-S} is the predecessor of \texttt{ada2} with the same number of parameters; results are copied from \citet{neelakantan2022text}.}
\label{table:beir}
\end{table*}

\subsection{BEIR}

We first evaluate the generalization capabilities of embedding APIs across a variety of domains.
To this end, we measure their effectiveness on BEIR \cite{beir}, a heterogeneous evaluation benchmark intended to gauge the domain generalization of retrieval models.
BEIR consists of 18 retrieval datasets across 9 domains and \citet{beir} showed that BM25 is a strong baseline, surpassing most dense retrieval models.

We adopt the embedding API as a re-ranking component on top of BM25 retrieved results. 
Re-ranking is a more realistic scenario, compared to full ranking, because the number of documents to encode in re-ranking is commensurate with the number of test queries, which is orders of magnitude smaller than the collection size, usually comprising millions of documents.
Thus, re-ranking is more efficient and cheaper than full ranking. 

For the BM25 retrieval, we use Anserini~\cite{anserini} to index the corpora in BEIR and retrieve top-100 passages for each dataset.
Then, the queries and the retrieved passages are encoded using the embedding APIs.
We reorder the retrieval output based on the similarity between query embeddings and those of the passages.

In addition to BM25, our baselines include the following dense retrieval models:

\begin{itemize}[leftmargin=*]
    \item TASB~\cite{hofstatter2021efficiently}, a prominent dense retrieval model that leverages topic-aware sampling of queries during training to construct more informative and more balanced contrastive examples in terms of sample difficulty. TASB is built on DistilBERT~\cite{distilbert} and is fine-tuned on MS MARCO~\cite{msmarco}.
    \item cpt~\cite{neelakantan2022text}, an earlier version of OpenAI's embedding service.
\end{itemize}

\noindent The results are presented in Table~\ref{table:beir}.\footnote{We did not test {\aalph}'s \texttt{luminous} on BEIR due to budget constraints.}
TASB re-ranking results show a $+$4\% increase over TASB full-ranking on average, showing that re-ranking via bi-encoder models is indeed a viable method.
We observe that {\openai}'s \texttt{ada2} is the most effective model, surpassing TASB and {\colarge} by $+$4.7\% and $+$2.4\% on average, respectively.
However, {\colarge} outperforms \texttt{ada2} on 5 tasks.
Specifically, {\colarge} achieves the highest nDCG@10 on NQ (question answering), SCIDOCS (citation prediction), Climate-FEVER (fact verification), and both duplicate question retrieval tasks, i.e., CQADupStack, and Quora.
Also, we observe that {\cosmall} trails {\colarge} by 2.6\% on average and is nearly on par with TASB. 

Finally, an interesting observation is that BM25 leads all other models on 3 tasks:\ BioASQ, Signal-1M, and Touch{\'e}-2020.
These are datasets collected based on lexical matching, suggesting that embedding APIs struggle in finding lexical overlaps.

\begin{table*}[t]
\centering
\small
\resizebox{\textwidth}{!}{
\begin{tabular}{lrccccccc}
\toprule
 & & \multicolumn{5}{c}{\textit{Full-ranking}}& \multicolumn{2}{c}{\textit{BM25 Top-100 Re-rank}} \\
\multicolumn{1}{c}{\textbf{Language}}& \multicolumn{1}{c}{\textbf{\# Q}}& \textbf{BM25} & \textbf{mDPR} & \textbf{mDPR+BM25} & \textbf{{\cohere}} & \small{\textbf{{\cohere}+BM25}} & \textbf{{\cohere}} & \textbf{luminous}$_\text{\texttt{base}}$\\
\cmidrule(lr){1-2} \cmidrule(lr){3-7} \cmidrule(lr){8-9} 
\Ar & 2,896& 0.481 & 0.499 & 0.673 & 0.617 & \textbf{0.686} & 0.667 & -\\
\Bn & 411& 0.508 & 0.443 & 0.654 & 0.594 & \textbf{0.676} & 0.634 & -\\
\De & 305& 0.226 & 0.490 & \textbf{0.565} & 0.436 & 0.468 & 0.414 & 0.396\\
\Es & 648& 0.319& 0.478 & \textbf{0.641} & 0.233 & 0.349 & 0.507 & 0.482\\
\Fa & 632& 0.333& 0.480 & \textbf{0.594} & 0.471 & 0.520 & 0.484 & -\\
\Fi & 1,271& 0.551& 0.472 & 0.672 & 0.634 & \textbf{0.716} & 0.675 & -\\
\Fr & 343& 0.183& 0.435 & \textbf{0.523} & 0.462 & 0.434 & 0.443 & 0.415\\
\Hi & 350& 0.458& 0.383 & 0.616 & 0.493 & \textbf{0.623} & 0.573 & -\\
\Id & 960& 0.449& 0.272 & 0.443 & 0.446 & \textbf{0.565} & 0.505 & -\\
\Ja & 860& 0.369& 0.439 & \textbf{0.576} & 0.460 & 0.557 & 0.516 & -\\
\Ko & 213& 0.419& 0.419 & \textbf{0.609} & 0.496 & 0.597 & 0.546 & -\\
\Ru & 1,252& 0.334& 0.407 & \textbf{0.532} & 0.469 & 0.528 & 0.447 & -\\
\Sw & 482& 0.383& 0.299 & 0.446 & \textbf{0.611} & 0.608 & 0.543 & -\\
\Te & 828& 0.494& 0.356 & 0.602 & 0.613 & \textbf{0.686} & 0.638 & -\\
\Th & 733& 0.484& 0.358 & 0.599 & 0.546 & \textbf{0.678} & 0.606 & -\\
\Yo & 119& 0.019& 0.396& 0.374 & \textbf{0.762} & 0.735 & 0.629 & -\\
\Zh & 393& 0.180& 0.512& \textbf{0.526} & 0.365 & 0.416 & 0.389 & -\\
\cmidrule(lr){1-2} \cmidrule(lr){3-7} \cmidrule(lr){8-9} 
\multicolumn{2}{c}{Avg. nDCG@10}& 0.364 & 0.420 & 0.567 & 0.512 & \textbf{0.579} & 0.542 & - \\
\bottomrule
\end{tabular}
}
\caption{
Results (nDCG@10) on the {\miracl} dev set across 17 languages for the full-ranking and re-ranking experiments. \textbf{\#~Q} indicates the number of queries in the dev set. Luminous only supports \De, \Es, and \Fr.
}
\label{tab:miracl}
\end{table*}

\subsection{Multilingual Retrieval: MIRACL}

We further assess the embedding APIs in the multilingual retrieval setting, where the aim is to build retrieval models that can operate in several languages while maintaining their retrieval effectiveness across languages.
For this purpose, we use MIRACL~\cite{miracl}, a large-scale multilingual retrieval benchmark that spans 18 languages with more than 725K relevance judgments collected from native speakers.

\begin{figure*}[t!]
     \centering
     \begin{subfigure}[b]{0.49\textwidth}
         \centering
         \includegraphics[width=\textwidth]{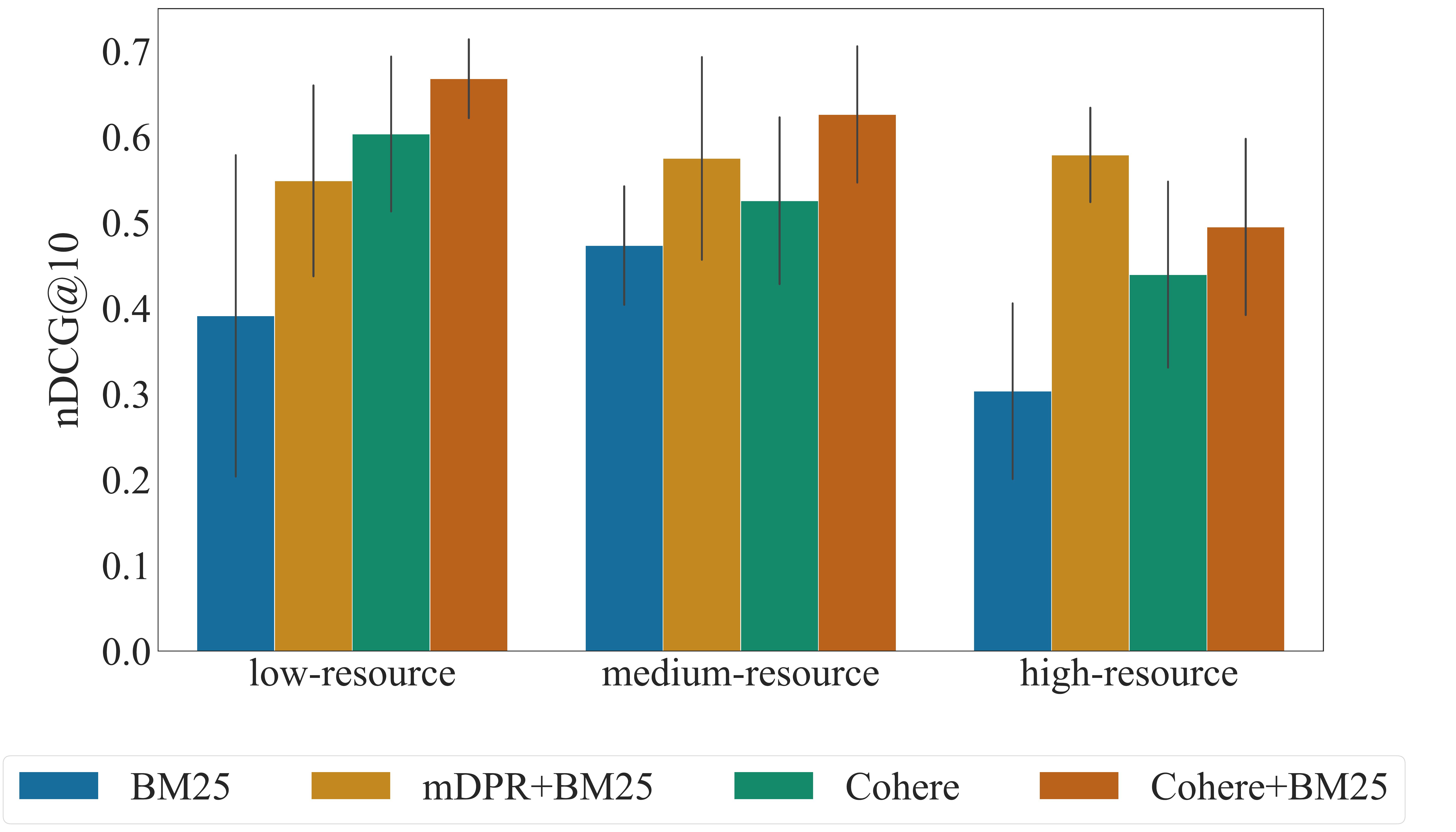}
         \label{fig:miracl-resource:ndcg}
     \end{subfigure}
     \hfill
    \begin{subfigure}[b]{0.49\textwidth}
         \centering
         \includegraphics[width=\textwidth]{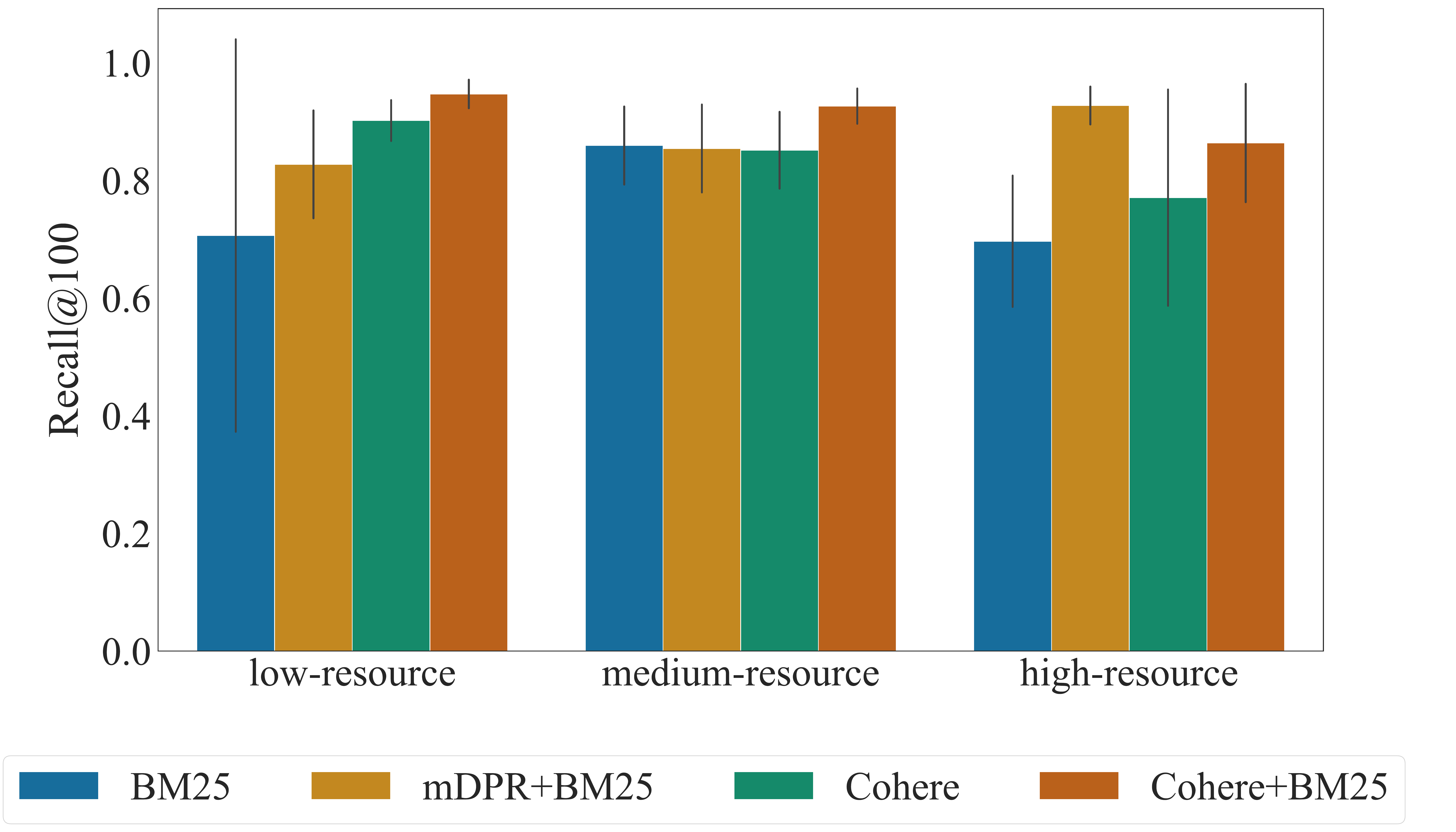}
         \label{fig:miracl-resource:recall}
     \end{subfigure}
    \caption{Average nDCG@10 (left) and Average Recall@100 (right) of full-ranking models on the MIRACL dev set for different categories of languages in terms of their available resources:\ low (\Bn, \Hi, \Sw, \Te, \Th, and \Yo), medium (\Fi, \Id, and \Ko), and high (\Ar, \De, \Es, \Fa, \Fr, \Ja, \Ru, and \Zh). The error bars show the standard deviation of nDCG@10 and Recall@100.}
    \label{fig:miracl-resource}
\end{figure*}

We test Cohere's multilingual model as well as {\aalph}'s \texttt{luminous} on MIRACL.
OpenAI does not recommend using their embeddings service for non-English documents and thus their API was omitted from this experiment.
Analogous to the previous experiment, we adopt a re-ranking strategy on top-100 passages retrieved by BM25.
For Cohere, we carry out full-ranking retrieval to draw a comparison with first-stage retrieval models.
We also construct a hybrid model combining BM25 and Cohere by interpolating their normalized retrieval scores, following \citet{miracl}.
The baselines are also taken from that paper:\ BM25, mDPR, and the hybrid model mDPR+BM25. We reuse the indexes provided in Pyserini~\cite{pyserini} to generate the baseline runs.
For all models, we measure nDCG@10 and Recall@100.

The results on the MIRACL dev set are reported in Table~\ref{tab:miracl}.
Re-ranking BM25 via {\cohere} yields better overall results (0.542), compared to full-ranking (0.512), which is consistent with our observation on BEIR.
However, while the two re-ranking models,  \texttt{luminous} and {\cohere}, surpass BM25 on all languages, they lag behind the full-ranking hybrid models.
The results show that the winning recipe here is to build a hybrid model, i.e., first perform retrieval on the entire corpus and then combine the results with BM25.
In particular, {\cohere}+BM25 achieves the highest average nDCG@10, outperforming the other models on 7 languages.
The second best model overall is the other hybrid model, mDPR+BM25, trailing Cohere+BM25 by $-$1.2\%.

We further investigate how the models perform on low-, medium-, and high-resource languages.
To this end, following the categorization of \citet{wu-dredze-2020-languages}, we group languages into three categories based on the number of articles they contain in Wikipedia, reported in \citet{miracl}:\ low-resource ($<$200K), medium-resource ($>$200K but $<$600K), and high-resource ($>$600K).
We measure the average nDCG@10 and Recall@100 for each language category.
The results are visualized in Figure~\ref{fig:miracl-resource}.
The effectiveness of BM25 on low-resource languages is nearly on par with its effectiveness on high-resource languages. 
Interestingly, mDPR+BM25 consistently performs well across the three language categories.
On the other hand, {\cohere}'s hybrid and standalone models excel on low-resource languages and are competitive with mDPR+BM25 on medium-resource languages.
However, on high-resource languages, mDPR+BM25 outperforms {\cohere}'s hybrid model due in part to the prevalence of text in these languages during mBERT pre-training \cite{wu-dredze-2020-languages} in mDPR.

\section{Conclusion}

The incredible capabilities of Transformer-based language models at scale have attracted a handful of companies to offer access to their proprietary LLMs via APIs.
In this paper, we aim to qualitatively and quantitatively examine semantic embedding APIs that can be used for information retrieval. 
Our primary focus is to assess existing APIs for domain generalization and multilingual retrieval.
Our findings suggest that re-ranking BM25 results is a suitable and cost-effective option for English; on the BEIR benchmark, {\openai}$_\text{\texttt{ada2}}$ performs the best on average.
In multilingual settings, while re-ranking remains a viable technique, a hybrid approach produces the most favorable results.
We hope that our insights aid practitioners and researchers in selecting appropriate APIs based on their needs in this rapidly growing market.

\section*{Limitations}

Similar to other commercial products, embedding APIs are subject to changes that could potentially impact their effectiveness, pricing, and usability.
Thus, it is important to note that our findings are specific to the APIs accessed during January and February 2023.
Nevertheless, we believe our evaluation framework can serve to thoroughly assess future releases of these APIs.

Moreover, we limit our focus to the effectiveness and robustness of semantic embedding APIs.
Nonetheless, safe deployment of retrieval systems for real-world applications necessitates the evaluation of their fairness as well as additional considerations. 
Despite their scale, language models have been found to learn, and sometimes perpetuate societal biases and harmful stereotypes ingrained in the training corpus \cite{bender2021dangers}.
Consequently, it is crucial to assess potential biases in the embedding APIs with respect to protected and marginalized groups.
This paper does not delve into this aspect of API evaluation and further research is required to examine these and other issues in real-world applications.

\section*{Acknowledgements}

We thank Aleph-Alpha for providing us with credits to explore and test their APIs.
This research was supported in part by the Natural Sciences and Engineering Research Council (NSERC) of Canada.

\bibliography{anthology,main}
\bibliographystyle{acl_natbib}

\end{document}